\documentclass[11pt]{article}

\usepackage[utf8]{inputenc}
\usepackage[T1]{fontenc}
\usepackage[margin=1in]{geometry}
\usepackage{amsmath,amssymb,amsthm}
\usepackage{mathtools}
\usepackage{graphicx}
\usepackage{booktabs}
\usepackage[dvipsnames]{xcolor}
\usepackage{algorithm}
\usepackage{subcaption}
\usepackage{algpseudocode}

\usepackage{placeins}

\usepackage[numbers,sort&compress]{natbib}

\usepackage[colorlinks=true,citecolor=blue,linkcolor=blue,urlcolor=blue]{hyperref}
\usepackage{doi}
\makeatletter
\providecommand{\theHALG@line}{\theHalgorithm.\arabic{ALG@line}}
\makeatother
\usepackage[capitalise]{cleveref}
\usepackage{titlesec}
\usepackage{fancyhdr}
\usepackage{enumitem}
\usepackage{siunitx}
\usepackage{orcidlink}

\titleformat{\section}{\normalfont\Large\bfseries}{\thesection}{1em}{}
\titleformat{\subsection}{\normalfont\large\bfseries}{\thesubsection}{1em}{}

\hypersetup{
	pdftitle={\title{\textbf{Electronic Origin of Ionic-Conductivity Suppression in Gd/W Co-doped La$_2$Mo$_2$O$_9$ from First-Principles Calculations}}},
	pdfauthor={Amogh U. Lanjewar, Saurabh Shiwankar, Smita Acharya},
	pdfsubject={Condensed Matter > Materials Science},
	pdfkeywords={La2Mo2O9, LAMOX, Gd/W co-doping, DFT+U, projected density of states, oxide-ion conductors, SOFC electrolytes}
}

\title{\textbf{Gd-$4f$ Exchange Splitting and Mo-$4d$ Crystal-Field Redistribution in Gd/W Co-doped La$_2$Mo$_2$O$_9$: A DFT+$U$ Study}}

\author{
	Amogh U. Lanjewar\orcidlink{0009-0009-0868-5781}, Saurabh Shiwankar\orcidlink{0009-0004-4121-5400}, Smita Acharya$^{*}$\orcidlink{0000-0001-8150-3336}\\
	Advanced Materials Research Laboratory (AMRL),\\
	Department of Physics,\\
	Rashtrasant Tukadoji Maharaj Nagpur University,\\
	Nagpur 440033, Maharashtra, India\\
\small $^{*}$Corresponding author:
\href{mailto:saha275@nagpuruniversity.ac.in }{saha275@nagpuruniversity.ac.in}
}

\date{August 2026}

\begin{document}
	\maketitle
	\thispagestyle{fancy}
	
\begin{abstract}
	La$_2$Mo$_2$O$_9$ (LAMOX) is a promising oxide-ion conductor for intermediate-temperature solid oxide fuel cells, but its practical application is limited by a first-order monoclinic-to-cubic ($\alpha\rightarrow\beta$) phase transition. Here, we investigate the electronic structure of pristine La$_2$Mo$_2$O$_9$ and the Gd/W co-doped composition La$_{1.6}$Gd$_{0.4}$Mo$_{1.7}$W$_{0.3}$O$_9$ using spin-polarized density functional theory with an on-site Hubbard correction for the localized Gd-$4f$ states. The projected density of states reveals that pristine La$_2$Mo$_2$O$_9$ is an O-$2p$/Mo-$4d$ charge-transfer oxide in which La contributes negligibly near the band edges. Gd/W co-doping introduces a strongly exchange-split Gd-$4f$ manifold with a majority--minority separation of approximately 10--11~eV, substantially redistributes the Mo-$4d$ electronic states through a threefold increase in crystal-field splitting, and reduces the O-$2p$ contribution near the valence-band maximum from 78\% to 66\%. These electronic signatures are consistent with the experimentally observed lattice contraction, Mo--O Raman-mode softening, and the non-monotonic evolution of oxide-ion conductivity across the co-doped series. In particular, the pronounced crystal-field stiffening and localized Gd magnetism at the highest doping level provide a microscopic electronic explanation for the observed suppression of ionic conductivity. The present results establish an atomistic electronic-structure framework for understanding dopant-induced phase stabilization and oxide-ion transport in Gd/W co-doped LAMOX electrolytes, providing design principles for improved solid oxide fuel cell materials.
\end{abstract}
	
	
\section{Introduction}
\label{sec:theory}

Solid oxide fuel cells (SOFCs) convert chemical energy directly into electricity with high efficiency and fuel flexibility, but the yttria-stabilised zirconia electrolytes used in first-generation devices require operating temperatures above $900\,^{\circ}$C, which limits their long-term durability and raises materials and balance-of-plant costs \cite{Jacobson2010,Gao2016}. Reducing the operating temperature into the intermediate range ($500$--$700\,^{\circ}$C) without sacrificing ionic conductivity has therefore become a central objective of solid-state electrolyte research, and has motivated a sustained search for oxide-ion conductors whose transport properties do not rely on a fluorite-type defect structure \cite{Malavasi2010,Yang2023}.

Among the alternative families identified for this purpose, the LAMOX materials based on La$_2$Mo$_2$O$_9$ attracted particular attention following the report by Lacorre \textit{et al.} of an oxide-ion conductivity in the cubic $\beta$ polymorph comparable to that of yttria-stabilised zirconia at substantially lower temperature \cite{Lacorre2000}. The parent compound undergoes a first-order structural transition near $580\,^{\circ}$C from a monoclinic, ordered $\alpha$ phase (space group $P2_1$) to a cubic, oxygen-disordered $\beta$ phase (space group $P2_13$), and it is the intrinsic oxygen sublattice disorder of the $\beta$ phase, rather than a change in the underlying cation framework, that is responsible for the enhanced ionic mobility; pair-distribution-function analysis has shown that the local coordination environment on either side of the transition is nearly indistinguishable, so that $\alpha\to\beta$ is more accurately described as an order-disorder transition of the oxygen sublattice than as a reconstructive phase change \cite{Malavasi2007Nature}. This transition, however, is also the principal obstacle to practical application: it is accompanied by an abrupt jump in conductivity and by anisotropic volume changes that can compromise mechanical and chemical stability over repeated thermal cycling \cite{Lacorre2006}.

Chemical substitution on either the La or the Mo site has been the primary strategy for suppressing the $\alpha\to\beta$ transition and stabilising the cubic, fast-conducting framework down to room temperature. Bi$^{3+}$ substitution for La$^{3+}$ has been shown, by temperature-controlled neutron diffraction, to enhance conductivity through distortion of the anti-tetrahedral MoO$_x$ framework and thermally activated escape of oxide ions toward partially occupied interstitial sites \cite{Corbel2011Structural}, while Eu$^{3+}$ substitution stabilises the cubic phase above a threshold concentration but introduces regions of pronounced thermal-history-dependent metastability at intermediate doping levels \cite{Corbel2007Topological}. Quasi-elastic neutron scattering combined with \textit{ab initio} molecular dynamics has further resolved the microscopic oxide-ion migration pathway in La$_2$Mo$_2$O$_9$ as a sequence of discrete jumps within and between flexible Mo-centred coordination polyhedra \cite{Peet2017Direct}, underlining that conductivity in this family is controlled less by the identity of the A-site cation than by the flexibility and connectivity of the Mo--O polyhedral network that the dopant leaves behind. Not every substitution is beneficial, however, and the conductivity response to doping is frequently non-monotonic and composition-dependent, which has motivated a shift in the field from single-site to co-doping strategies that simultaneously target the La and Mo sublattices \cite{Lacorre2006}.

A complementary, electronic-structure perspective on these substitution effects has emerged more slowly. Density-functional theory (DFT) calculations of pristine La$_2$Mo$_2$O$_9$ show a valence band dominated by O-2$p$ states and a conduction band dominated by unoccupied Mo-4$d$ states, with La-derived states pushed to higher energy away from both band edges, so that the compound behaves electronically as an O-to-Mo charge-transfer oxide with La acting largely as a structural, rather than electronic, spectator \cite{Colmont2020Origin}. This picture is consistent with the broader conclusion, drawn from first-principles studies of oxide-ion conductors more generally, that migration barriers and vacancy energetics are governed primarily by the transition-metal--oxygen sublattice, while A-site cations modulate these properties indirectly through lattice strain and polarizability \cite{MunozGarcia2014}. Extending this electronic-structure treatment to open-shell rare-earth dopants, however, introduces an additional complication: partially filled $4f$ shells are poorly described by semi-local exchange-correlation functionals because of uncorrected self-interaction error, which spuriously delocalises the $f$ manifold and places it close to the Fermi level; an on-site Hubbard correction is required to recover the correct, strongly localised character of such states \cite{Anisimov1991,Cococcioni2005}.

Gadolinium is a particularly instructive dopant in this context. As Gd$^{3+}$, it carries a half-filled $4f^7$ ($^8S_{7/2}$) configuration with one of the largest magnetic moments among the lanthanides, so that its incorporation onto the La site of La$_2$Mo$_2$O$_9$ introduces both an isovalent structural perturbation, expected to leave the oxygen-vacancy chemistry largely unchanged, and a genuinely new, localised magnetic degree of freedom absent from the non-magnetic parent compound. A concurrent experimental study of the co-doped series La$_{2-x}$Gd$_x$Mo$_{1.7}$W$_{0.3}$O$_{9-\delta}$ has shown that combined Gd/W substitution suppresses the $\alpha\to\beta$ transition across the entire composition range, stabilising the cubic $P2_13$ framework at room temperature, while the bulk ionic conductivity varies non-monotonically with Gd content and reaches a maximum at intermediate doping before collapsing at the highest substitution level studied \cite{Shiwankar2025}. This behaviour indicates that Gd/W co-doping modifies the LAMOX framework through a more complex mechanism than simple isovalent dilution, and it raises the question of how the accompanying electronic and, potentially, magnetic reorganisation of the Mo/W--O and Gd sublattices is reflected at the level of the underlying electronic structure, a question that has not, to our knowledge, previously been addressed computationally for this specific co-doped composition.

Answering this question requires not only an accurate treatment of the Gd-4$f$ electrons, but also a supercell large enough to resolve the exchange interactions that a magnetically active Mo (and, prospectively, Gd) sublattice can support. Extracting reliable exchange constants from collinear-configuration total-energy mapping is highly sensitive to supercell size and shape: coordination shells that fold onto one another under periodic boundary conditions produce linearly dependent mapping equations from which individual exchange constants cannot be resolved, so that the farthest interaction shell accessible in a given calculation is a property of the specific supercell chosen rather than of its volume alone \cite{Sadeghi2015}. Systematic, symmetry-aware enumeration of candidate supercells, of the kind made possible by algorithms for generating derivative superstructures \cite{Hart2008} and implemented for the specific purpose of optimising Heisenberg-exchange supercells in the \texttt{SUPERHEX} methodology \cite{Alaei2025}, therefore provides a principled alternative to the arbitrary, uniformly enlarged cells conventionally used for this purpose.

In this work, we combine these two elements, a symmetry-optimised supercell search and a Hubbard-corrected first-principles electronic-structure treatment, to examine how Gd/W co-doping reorganises the electronic structure of La$_2$Mo$_2$O$_9$ relative to the pristine compound. Section~\ref{sec:method} describes the construction of the exchange-interaction supercells used for both materials and the details of the spin-polarised DFT+U calculations performed on them. Section~\ref{sec:results} presents the resulting orbital-resolved projected density of states for the pristine and Gd/W co-doped end-members and correlates the computed structural and electronic-structure trends directly with the concurrent experimental structural, vibrational and transport data of Shiwankar \textit{et al.} \cite{Shiwankar2025}, situating the present first-principles results within the wider body of structural and mechanistic work on doped La$_2$Mo$_2$O$_9$ \cite{Lacorre2000,Malavasi2007Nature,Corbel2011Structural,Corbel2007Topological,Peet2017Direct,Lacorre2006,Colmont2020Origin}.

\section{METHOD}
\label{sec:method}
\subsection{Building the Supercell}
\label{sec:building-supercell}

Extraction of Heisenberg exchange parameters $J_i$ from total-energy differences between collinear magnetic configurations requires a supercell large enough that each coordination shell around a magnetic site can be probed independently. In a cell that is too small, periodic images fold neighbouring shells back onto one another, so that the corresponding energy-mapping coefficients become linearly dependent and the associated $J_i$ cannot be resolved \cite{Alaei2025}. Rather than enlarging the parent cell uniformly by an arbitrary integer factor, which is computationally wasteful, we searched for the smallest and most efficient supercell of each parent structure using the \texttt{SUPERHEX} package \cite{Alaei2025,SuperhexGitHub}, following the linear-algebra formalism described below.

For a set of $n$ collinear magnetic configurations built on a common supercell, the total energy $E_k$ of the $k$-th configuration is related to the exchange constants through the mapping equation

\begin{equation}
	E_k = \sum_{i=1}^{m} \alpha_{k,i}\, J_i + c_0 ,
	\label{eq:mapping}
\end{equation}

\noindent where $\alpha_{k,i} \in \{+1,-1\}$ encodes the relative alignment of spin pairs separated by the $i$-th coordination shell in configuration $k$, $J_i$ is the corresponding exchange constant, and $c_0$ is a configuration-independent offset \cite{Sadeghi2015,Alaei2025}. Collecting Eq.~\eqref{eq:mapping} for $n$ configurations gives the linear system

\begin{equation}
	\mathbf{A}\,\mathbf{J} = \mathbf{E}, \qquad
	\mathbf{A} =
	\begin{pmatrix}
		1 & \alpha_{1,1} & \cdots & \alpha_{1,m} \\
		1 & \alpha_{2,1} & \cdots & \alpha_{2,m} \\
		\vdots & \vdots & \ddots & \vdots \\
		1 & \alpha_{n,1} & \cdots & \alpha_{n,m}
	\end{pmatrix},
	\label{eq:linsys}
\end{equation}

\noindent with $\mathbf{J} = (c_0, J_1, \dots, J_m)^{\mathsf{T}}$ and $\mathbf{E} = (E_1,\dots,E_n)^{\mathsf{T}}$. Reducing $\mathbf{A}$ to row-echelon form identifies the first column that is linearly dependent on the preceding ones; if this occurs at shell $q+1$, exchange constants beyond $J_q$ cannot be determined reliably in that supercell, and $J_{i \geq q}$ must be set to zero to avoid spurious values \cite{Alaei2025}. The farthest resolvable shell therefore depends on the size \emph{and} the shape of the supercell, not on its volume alone, which motivates a systematic search over candidate cells rather than a fixed conventional expansion.

Candidate supercells are constructed from the parent lattice vectors $\mathbf{V}_p$ through an integer transformation $\mathbf{V}_s = \mathbf{H}\mathbf{V}_p$, where $\mathbf{H}$ is a Hermite normal form (HNF) matrix,

\begin{equation}
	\mathbf{H} =
	\begin{pmatrix}
		a & b & c \\
		0 & d & e \\
		0 & 0 & f
	\end{pmatrix},
	\qquad 0 \leq b < d,\ \ 0 \leq c,e < f ,
	\label{eq:hnf}
\end{equation}

\noindent with integer entries constrained by $|\mathbf{H}| = a\,d\,f = m$, the target supercell volume relative to the parent cell \cite{Santoro1973,Santoro1972,Alaei2025}. Enumerating all matrices satisfying Eq.~\eqref{eq:hnf} for a given $m$, and removing duplicates related by the point-group symmetry of the parent lattice, yields every symmetry-inequivalent supercell of that volume \cite{Hart2008,Hart2009}. Because HNF transformations alone can produce highly skewed, elongated cells, each candidate is subsequently Minkowski-reduced to obtain a more compact, near-orthogonal representation without changing its volume \cite{NguyenStehle2009}. For every reduced candidate, a set of randomly generated collinear magnetic configurations on the Mo sublattice is used to build the coefficient matrix of Eq.~\eqref{eq:linsys}; duplicate rows arising from residual symmetry are discarded, and the rank and first dependent column of $\mathbf{A}$ are recorded together with the fraction of configurations that remain linearly independent.

We applied this procedure separately to the 144-atom parent cells of La$_2$Mo$_2$O$_9$ (LMO) and La$_{1.6}$Gd$_{0.4}$Mo$_{1.7}$W$_{0.3}$O$_9$ (LMOG4), designating Mo as the magnetic sublattice and using a real-space interaction cutoff of \SI{25}{\angstrom}. Supercell volumes were scanned over the range $m = 1$--$16$ relative to each parent cell, with $200$ random magnetic configurations generated per candidate ($\texttt{seed} = 42$) to test the independence of the coefficient matrix. For each material this produced a ranked list of symmetry-distinct supercells, tabulated by volume $m$, structure index $n$, matrix rank, index of the first dependent column, farthest permitted exchange shell $J_{\text{max}}$, percentage of independent magnetic configurations, and the variance of the reduced lattice vector lengths, the last of which serves as a practical measure of cell compactness. Supercell files are written as \texttt{cell-vol}$m$\texttt{-num}$n$\texttt{.vasp}, allowing direct cross-reference with the corresponding row of the analysis table.

Selecting among the resulting candidates involves a trade-off: larger volumes generally permit longer-range exchange interactions to be resolved, but the cost of the underlying spin-polarised DFT calculations grows steeply with system size for both materials, whose parent cells already contain 144 atoms. Balancing the range of accessible exchange shells against the computational resources available for the LMOG4 supercell, we selected the \texttt{cell-vol4-num9} structure for both LMO and LMOG4, which offers a favourable combination of a high percentage of independent configurations, a compact lattice-vector distribution, and a permitted exchange range sufficient to capture the dominant near-neighbour Mo--Mo interactions without requiring an intractably large supercell. The primitive and selected supercell representations of LMO and LMOG4 are compared in Fig.~\ref{fig:supercell-compare}.

\vspace{-0.5em}
\begin{figure*}[!htbp]
	\centering
	
	\begin{subfigure}[t]{0.48\textwidth}
		\centering
		\includegraphics[width=0.97\linewidth]{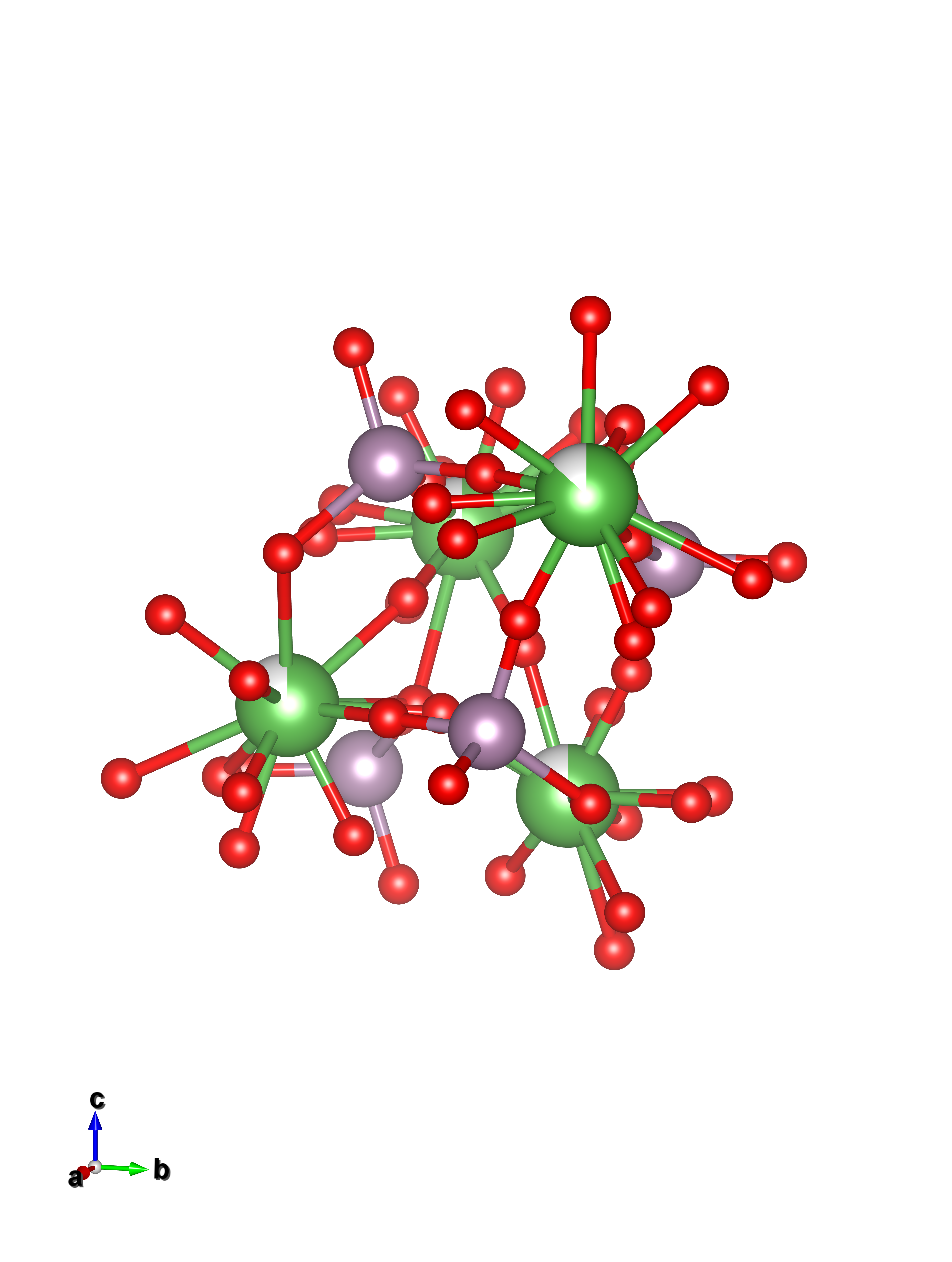}
		\caption{Primitive unit cell of LMO.}
		\label{fig:lmo_unit}
	\end{subfigure}
	\hfill
	\begin{subfigure}[t]{0.48\textwidth}
		\centering
		\includegraphics[width=0.97\linewidth]{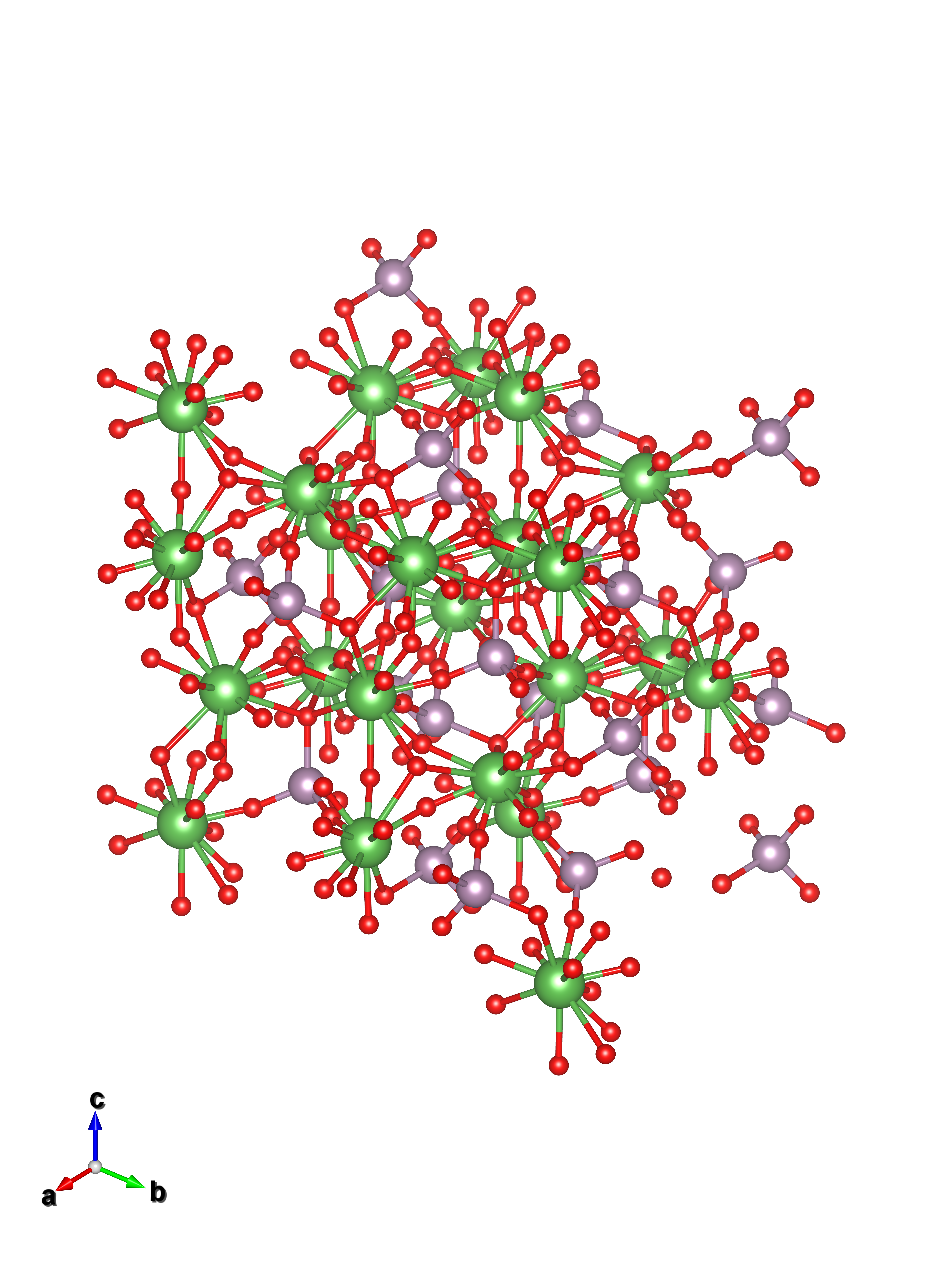}
		\caption{\texttt{cell-vol4-num9} magnetic supercell of LMO generated using \texttt{SUPERHEX}.}
		\label{fig:lmo_super}
	\end{subfigure}
	
	\vspace{0.8em}
	
	\begin{subfigure}[t]{0.48\textwidth}
		\centering
		\includegraphics[width=0.97\linewidth]{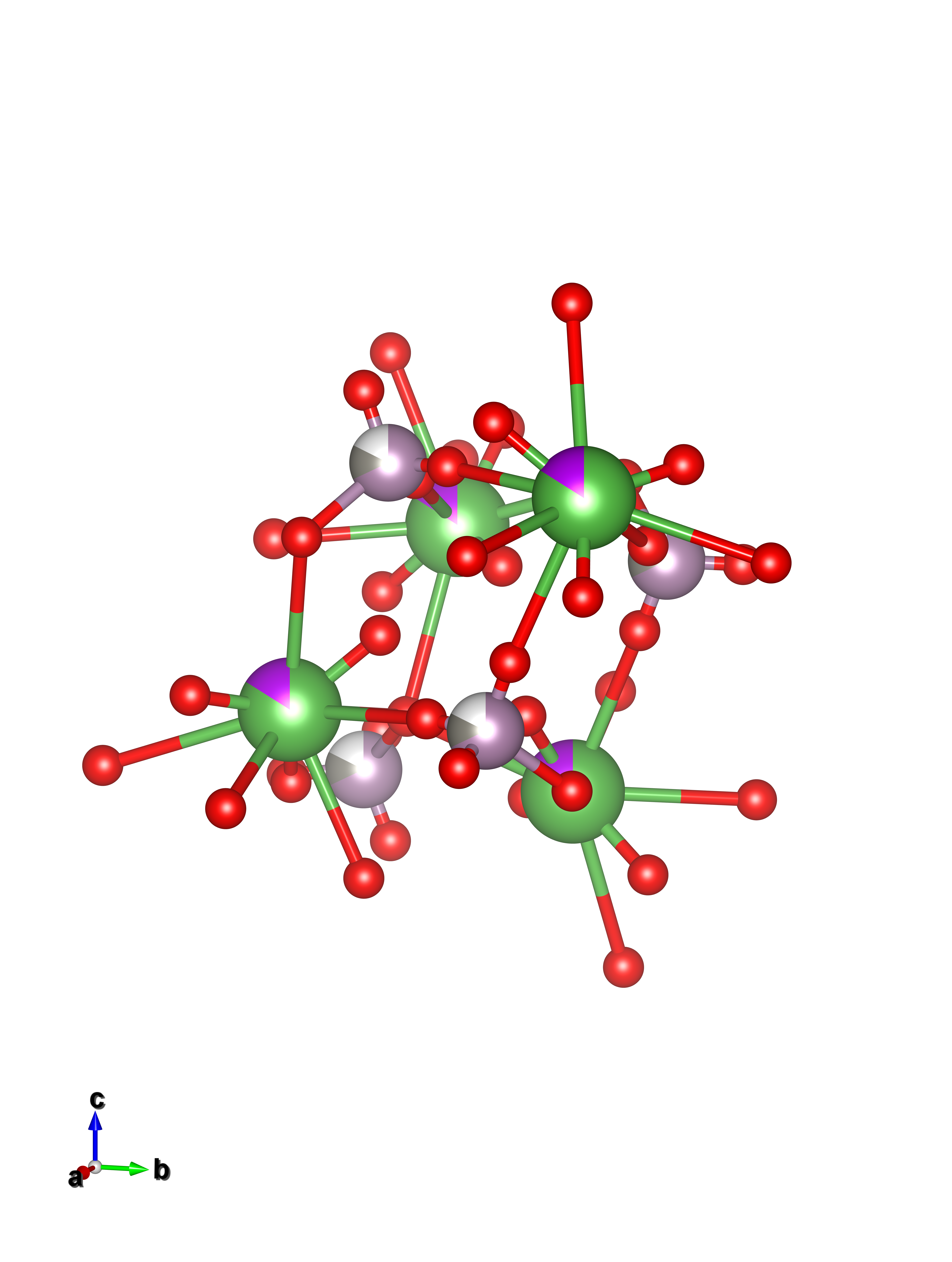}
		\caption{Primitive unit cell of LMOG4.}
		\label{fig:lmog4_unit}
	\end{subfigure}
	\hfill
	\begin{subfigure}[t]{0.48\textwidth}
		\centering
		\includegraphics[width=0.97\linewidth]{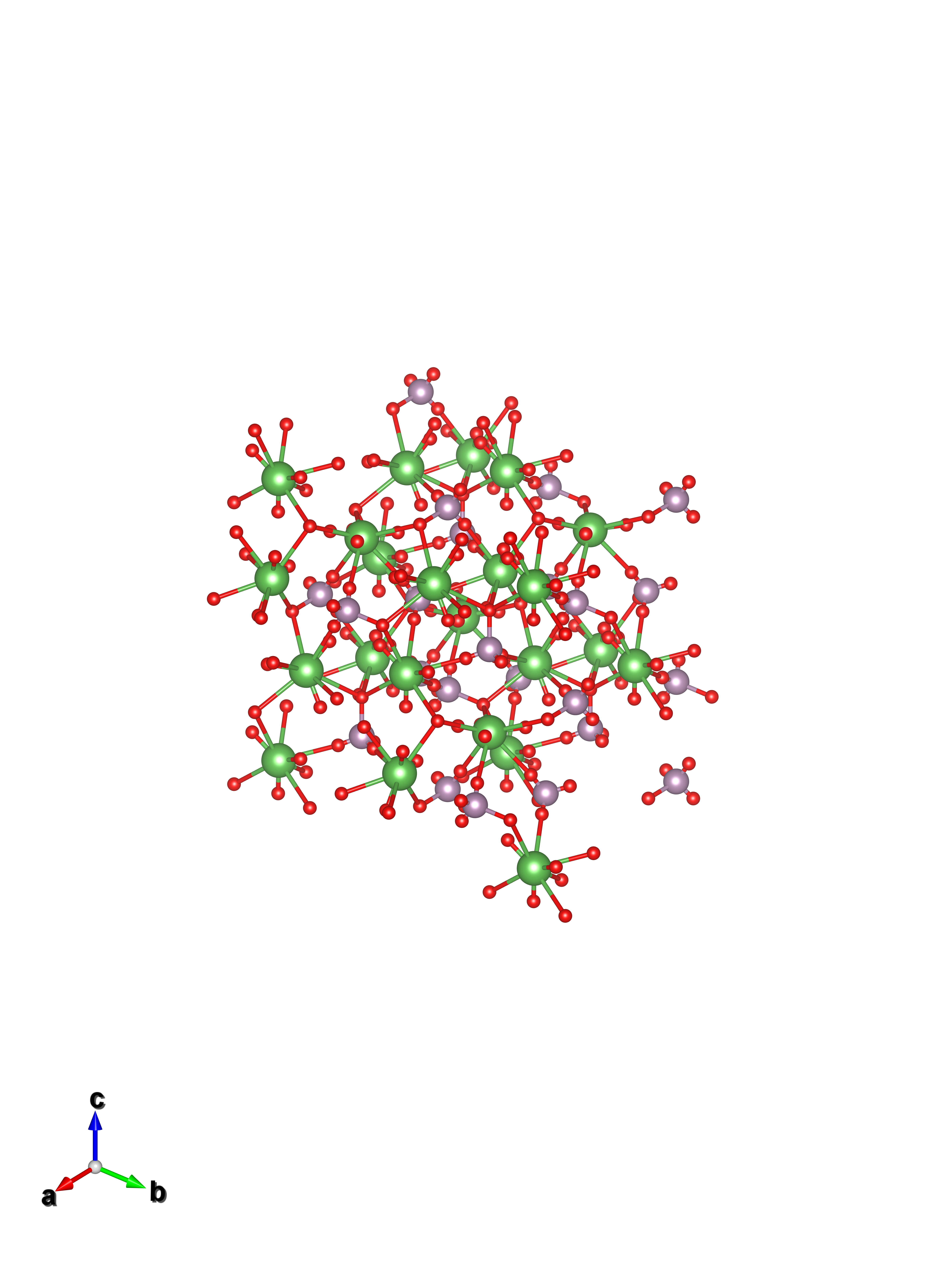}
		\caption{\texttt{cell-vol4-num9} magnetic supercell of LMOG4 generated using \texttt{SUPERHEX}.}
		\label{fig:lmog4_super}
	\end{subfigure}
	
	\caption{
		Crystal structures of (a) La$_2$Mo$_2$O$_9$ (LMO) and (c) La$_{1.6}$Gd$_{0.4}$Mo$_{1.7}$W$_{0.3}$O$_9$ (LMOG4), together with their corresponding
		\texttt{cell-vol4-num9} magnetic supercells ((b) and (d)) generated using
		\texttt{SUPERHEX}~\cite{Alaei2025} for the calculation of Mo--Mo Heisenberg
		exchange interactions.
		Atomic species are colour-coded:
		{\color{Green}\raisebox{0.2ex}{\rule{1.4ex}{1.4ex}}}\,La,
		{\color{Violet}\raisebox{0.2ex}{\rule{1.4ex}{1.4ex}}}\,Gd,
		{\color{lightgray}\raisebox{0.2ex}{\rule{1.4ex}{1.4ex}}}\,Mo,
		{\color{gray}\raisebox{0.2ex}{\rule{1.4ex}{1.4ex}}}\,W, and
		{\color{Red}\raisebox{0.2ex}{\rule{1.4ex}{1.4ex}}}\,O.
	}
	
	\label{fig:supercell_compare}
\end{figure*}

\subsection{DFT Implementation}
\label{sec:dft-implementation}

All total-energy calculations were carried out within the Kohn--Sham formulation of density functional theory (DFT), as implemented in the plane-wave, pseudopotential code \texttt{Quantum ESPRESSO} \cite{Giannozzi2009,Giannozzi2017}. Exchange and correlation were treated at the generalised-gradient approximation (GGA) level using the Perdew--Burke--Ernzerhof (PBE) functional \cite{Perdew1996} for both the pristine La$_2$Mo$_2$O$_9$ (LMO) and the Gd/W co-doped La$_{1.6}$Gd$_{0.4}$Mo$_{1.7}$W$_{0.3}$O$_9$ (LMOG4) cells.

Both structures were treated as 144-atom periodic cells, following directly from the LAMOX-type $\beta$-La$_2$Mo$_2$O$_9$ parent lattice. LMO was described in a rhombohedral setting (\texttt{ibrav}~=~5) with lattice parameter $a = 12.3809$~\r{A} and interaxial cosine $\cos(\alpha_{ab}) = -1/3$, while LMOG4, obtained after Gd/W substitution, was constructed as an explicit triclinic cell (\texttt{ibrav}~=~0) with lattice vectors of magnitude $\approx 7.1295$~\r{A}, entered directly via \texttt{CELL\_PARAMETERS}. Each cell contains 32 rare-earth (La, or La+Gd) sites, 16 Mo(+W) sites and 96 O sites, consistent with the nominal La$_2$Mo$_2$O$_9$ stoichiometry and its doped derivative.

Owing to the large size of the two supercells, reciprocal-space integration was restricted to the $\Gamma$ point only (a $1\times1\times1$ \texttt{automatic} $k$-mesh with zero shift). This choice reflects a direct computational-resource trade-off: for a 144-atom cell, a finite Monkhorst--Pack mesh \cite{MonkhorstPack1976} would multiply the cost of every self-consistent-field (SCF) cycle by the number of sampled $k$-points, whereas $\Gamma$-point sampling is a well-established and adequate approximation for large, low-symmetry supercells with a small reciprocal-cell volume \cite{Giannozzi2009}, and matches the sampling used to construct the exchange-interaction supercells described in Sec.~\ref{sec:building-supercell}.

Core--valence interactions were represented by pseudopotentials selected element-by-element from the \texttt{PSlibrary} and related distributions, with the pseudopotential family chosen for each species to balance transferability against plane-wave cost. For LMO, ultrasoft pseudopotentials \cite{Vanderbilt1990} were used throughout (Mo, O and La, all of the \texttt{rrkjus} PBE \texttt{psl.1.0.0} family), which permits a comparatively larger dataset (three species, no rare-earth $4f$ electrons other than La) to be run with a single, moderately expensive cutoff. LMOG4, containing five species including the open-shell $4f$ ion Gd$^{3+}$ and the heavier $5d$ element W, was instead treated with a mixed pseudopotential scheme: projector-augmented-wave (PAW) potentials \cite{Blochl1994,KresseJoubert1999} for Gd, La and O, an optimised norm-conserving Vanderbilt (ONCV) pseudopotential \cite{Hamann2013} for Mo, and an ultrasoft pseudopotential for W. This heterogeneous choice reflects pseudopotential availability and stability for each element within the PBE \texttt{PSlibrary} set \cite{DalCorso2014} and was found empirically to give a more robust SCF convergence for the doped, lower-symmetry cell than a single pseudopotential family.

The plane-wave kinetic-energy cutoff for the wavefunctions, $E_{\text{cut}}^{\text{wfc}}$, and for the augmented charge density, $E_{\text{cut}}^{\rho}$, satisfy

\begin{equation}
	|\mathbf{G}+\mathbf{k}|^2 \leq \frac{2m}{\hbar^2}E_{\text{cut}}^{\text{wfc}}, \qquad
	E_{\text{cut}}^{\rho} = \eta \, E_{\text{cut}}^{\text{wfc}},
	\label{eq:cutoffs}
\end{equation}

\noindent where $\eta$ is a species-dependent scaling factor required for an accurate representation of the augmentation charges of ultrasoft/PAW pseudopotentials. LMO was run with $E_{\text{cut}}^{\text{wfc}} = 115$~Ry and $E_{\text{cut}}^{\rho} = 1150$~Ry ($\eta = 10$), consistent with the harder ultrasoft Mo and La potentials used in that calculation, whereas LMOG4 was run with a reduced $E_{\text{cut}}^{\text{wfc}} = 85$~Ry and $E_{\text{cut}}^{\rho} = 850$~Ry (again $\eta = 10$). The lower cutoff for the doped cell was chosen deliberately: the ONCV Mo potential and the PAW La/Gd/O potentials converge at a softer cutoff than the \texttt{rrkjus} set, and reducing $E_{\text{cut}}^{\text{wfc}}$ from 115 to 85~Ry for the already larger and chemically more complex five-species system substantially reduced the memory and wall-time cost per SCF iteration without compromising the accuracy of the relative energetics, which was the resource-constrained priority for this system.

Partial occupations were introduced through first-order Methfessel--Paxton-type smearing of the Kohn--Sham eigenvalues rather than fixed integer occupations, since both structures are metallic or near-metallic in the GGA description (Sec.~\ref{sec:results}). For LMO, Fermi--Dirac smearing was used,

\begin{equation}
	f_i = \left[1 + \exp\!\left(\frac{\varepsilon_i - \varepsilon_F}{\sigma}\right)\right]^{-1}, \qquad \sigma = 0.01~\text{Ry},
	\label{eq:fermidirac}
\end{equation}

\noindent while LMOG4 used the Marzari--Vanderbilt "cold smearing" function \cite{MarzariVanderbilt1999},

\begin{equation}
	f_i = \frac{1}{2}\,\mathrm{erfc}\!\left(x_i + \frac{1}{\sqrt{2}}\right) + \frac{1}{\sqrt{2\pi}}\exp\!\left[-\left(x_i+\frac{1}{\sqrt{2}}\right)^2\right], \qquad x_i = \frac{\varepsilon_i-\varepsilon_F}{\sigma},
	\label{eq:coldsmearing}
\end{equation}

\noindent with a broadening $\sigma = 0.0125$~Ry. Cold smearing was preferred for LMOG4 because it minimises the spurious entropic contribution to the total energy while still stabilising the SCF cycle in the presence of the partially filled, more disordered Gd/W-derived states near the Fermi level; for the simpler LMO cell, Fermi--Dirac smearing at a slightly tighter width was sufficient and computationally cheaper to converge.

Both systems were treated within collinear spin-polarised DFT (\texttt{nspin}~=~2). Symmetry-breaking initial magnetic moments were assigned per species to seed the SCF cycle towards a magnetic solution: for LMO, starting magnetisations of $0.2\,\mu_B$ were placed on Mo and La, with O left non-magnetic; for LMOG4, a larger starting moment of $0.7\,\mu_B$ was placed on Gd, consistent with the half-filled $4f^7$ ($^8S_{7/2}$) configuration expected for Gd$^{3+}$, together with a smaller $0.2\,\mu_B$ seed on Mo. Because Gd $4f$ states are strongly localised and poorly described by semi-local GGA functionals, an on-site Hubbard correction \cite{Cococcioni2005} of the simplified rotationally invariant form,

\begin{equation}
	E_{U} = \frac{U}{2}\sum_{m,\sigma} n^{\sigma}_{m}\left(1 - n^{\sigma}_{m}\right),
	\label{eq:hubbard}
\end{equation}

\noindent was applied to the Gd $4f$ manifold in LMOG4 with $U = 6.0$~eV, using orthogonalised atomic (\texttt{ortho-atomic}) projectors \cite{Agapito2015} to define the occupation matrix $n^{\sigma}_{m}$. This DFT+U correction was applied only to LMOG4, since LMO contains no open-shell $4f$ or $3d$ ion requiring such treatment; no Hubbard correction was applied to Mo $4d$ or W $5d$ states in either cell.

Charge-density mixing used the modified Broyden scheme with a mixing parameter of 0.4 for LMO and 0.2 for LMOG4; the smaller mixing fraction adopted for the doped cell reflects the slower, less stable convergence typically encountered when mixing a Hubbard-corrected, multi-species magnetic system, and was selected empirically to avoid charge-sloshing instabilities without excessively increasing the number of SCF iterations required. For LMOG4, the starting potential and wavefunctions were constructed from a superposition of atomic charge densities and atomic orbitals with a random perturbation, which improves the likelihood of locating the correct magnetic ground state when several near-degenerate spin configurations are accessible.

Overall, the cutoffs, pseudopotential families, smearing schemes and mixing parameters for LMO and LMOG4 were not chosen identically, but were tuned independently for each system to obtain the most reliable SCF convergence at the lowest practical computational cost, given the constraint that a large number of distinct collinear magnetic configurations of a 144-atom cell had to be evaluated for the exchange-parameter extraction described in Sec.~\ref{sec:building-supercell}.

\section{RESULTS and DISCUSSIONS}
\label{sec:results}
\subsection{PDOS}
\label{sec:pdos}

The orbital character of the electronic states from the self-consistent calculations of Sec.~\ref{sec:dft-implementation} was resolved by projecting the plane-wave Kohn--Sham wavefunctions onto atomic-like orbitals, using the L\"owdin-orthogonalised scheme implemented in \texttt{projwfc.x} \cite{SanchezPortal1995}. For an atom $A$ and angular-momentum channel $l$, the spin-resolved projected density of states (PDOS) is

\begin{equation}
	\rho^{\sigma}_{A,l}(E) = \sum_{i,\mathbf{k},m} w_{\mathbf{k}}\left|\left\langle \phi_{A,l,m}\,\middle|\,\psi^{\sigma}_{i\mathbf{k}}\right\rangle\right|^2 \delta\!\left(E-\varepsilon^{\sigma}_{i\mathbf{k}}\right),
	\label{eq:pdos-def}
\end{equation}

\noindent where $\phi_{A,l,m}$ is the atomic projector, $\psi^{\sigma}_{i\mathbf{k}}$ the Kohn--Sham state of spin $\sigma$, band $i$, and $\mathbf{k}$-point weight $w_{\mathbf{k}}$. The sum over $i,\mathbf{k}$ is restricted to $\mathbf{k}=\Gamma$ for both supercells, as in Sec.~\ref{sec:dft-implementation}. Summing Eq.~\eqref{eq:pdos-def} over all atoms of a given element and orbital gives the orbital-resolved PDOS discussed below. The delta function was represented numerically on the DOS energy grid, and all energies are quoted relative to the Fermi level, $E-E_F$.

\subsubsection*{LMO: an O-2p/Mo-4d charge-transfer manifold}

\begin{figure}[!htbp]
	\centering
	\includegraphics[width=0.78\textwidth]{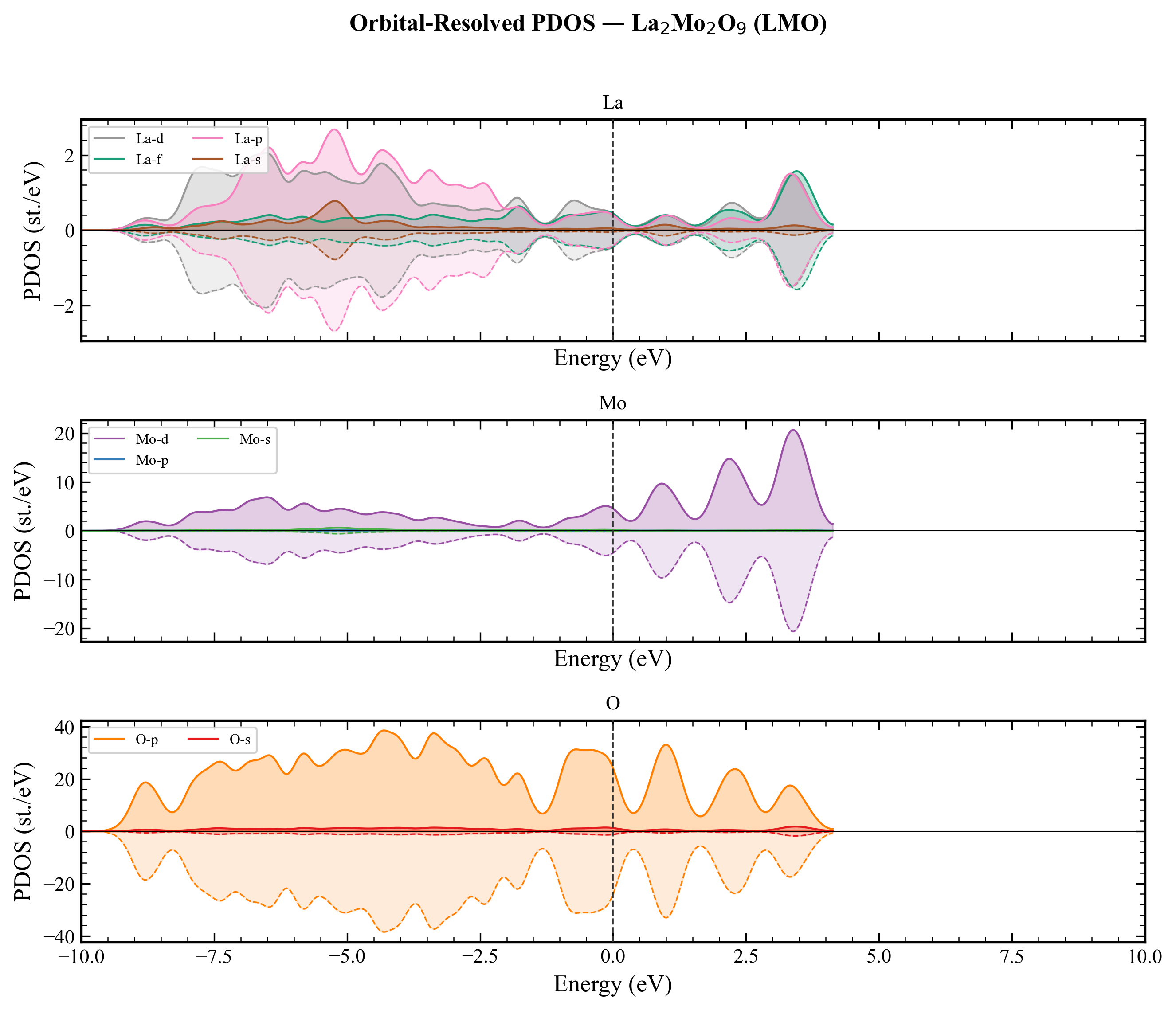}
	\caption{Orbital-resolved, spin-decomposed projected density of states of LMO (La$_2$Mo$_2$O$_9$), separated by element (La, Mo, O). Solid curves denote the spin-up channel and dashed curves the spin-down channel (plotted as negative PDOS for clarity); the vertical dashed line marks $E_F$.}
	\label{fig:lmo-pdos-stacked}
\end{figure}

Figure~\ref{fig:lmo-pdos-stacked} shows the element- and orbital-resolved PDOS of LMO (La, Mo, O), with spin-up and spin-down channels plotted as mirror images. The two channels overlap almost exactly across the full $-10$ to $+10$~eV window. This confirms the near-zero net spin polarisation already noted from the total-DOS analysis of Sec.~\ref{sec:dft-implementation} ($P \approx 0.0\%$). The valence band is O-2$p$ in character. The O panel of Fig.~\ref{fig:lmo-pdos-stacked} shows a dense run of peaks between $-9$ and $-1$~eV, reaching about 40 states/eV per spin channel, and tracking the total-DOS maxima at $E-E_F = -8.784,\,-7.374,\,-6.484,\,-5.814,\,-5.104,\,-4.324,\,-3.374$ and $-2.424$~eV. La-$d$ and La-$p$ states add only a weak background ($\lesssim 3$ states/eV) over the same range, and the Mo-$d$ shoulder is smaller still ($\lesssim 7$ states/eV). The top of the occupied manifold is, for practical purposes, an oxygen band. This agrees with the O-2$p$-dominated valence band reported for La$_2$Mo$_2$O$_9$ from hybrid-functional PDOS calculations \cite{Colmont2020Origin}, and with a broader pattern seen in first-principles studies of oxide-ion conductors: the A-site rare-earth cation tends to sit back structurally rather than contribute frontier orbitals \cite{MunozGarcia2014}.

The unoccupied states are, by contrast, dominated by Mo-4$d$. The Mo panel of Fig.~\ref{fig:lmo-pdos-stacked} shows three separated conduction-band peaks near $+1$, $+2$ and $+3.5$~eV, the tallest reaching $\sim$21 states/eV, an order of magnitude above the corresponding La- or O-derived weight. This multi-peak Mo-4$d$ structure matches the moderate octahedral crystal-field splitting, $\Delta_{\mathrm{CF}} \approx 1.18$~eV, extracted independently from the bimodal Mo-$d$ PDOS in Sec.~\ref{sec:dft-implementation}, and separates a lower $t_{2g}$-like feature from an upper $e_g$-like one. An O-2$p$ valence band next to a Mo-4$d$ conduction band, with low but non-zero density of states in between, is the charge-transfer-oxide picture already reported for La$_2$Mo$_2$O$_9$ \cite{Colmont2020Origin}. Our PBE-GGA calculation places finite weight directly at $E_F$ rather than opening the $\sim$3.7~eV gap seen with hybrid functionals or in experiment \cite{Colmont2020Origin}. This gap underestimation in $d$-electron oxides is a long-standing limitation of semi-local exchange-correlation functionals \cite{PerdewLevy1983}, not a peculiarity of this system.

\begin{figure}[!htbp]
	\centering
	\includegraphics[width=0.70\textwidth]{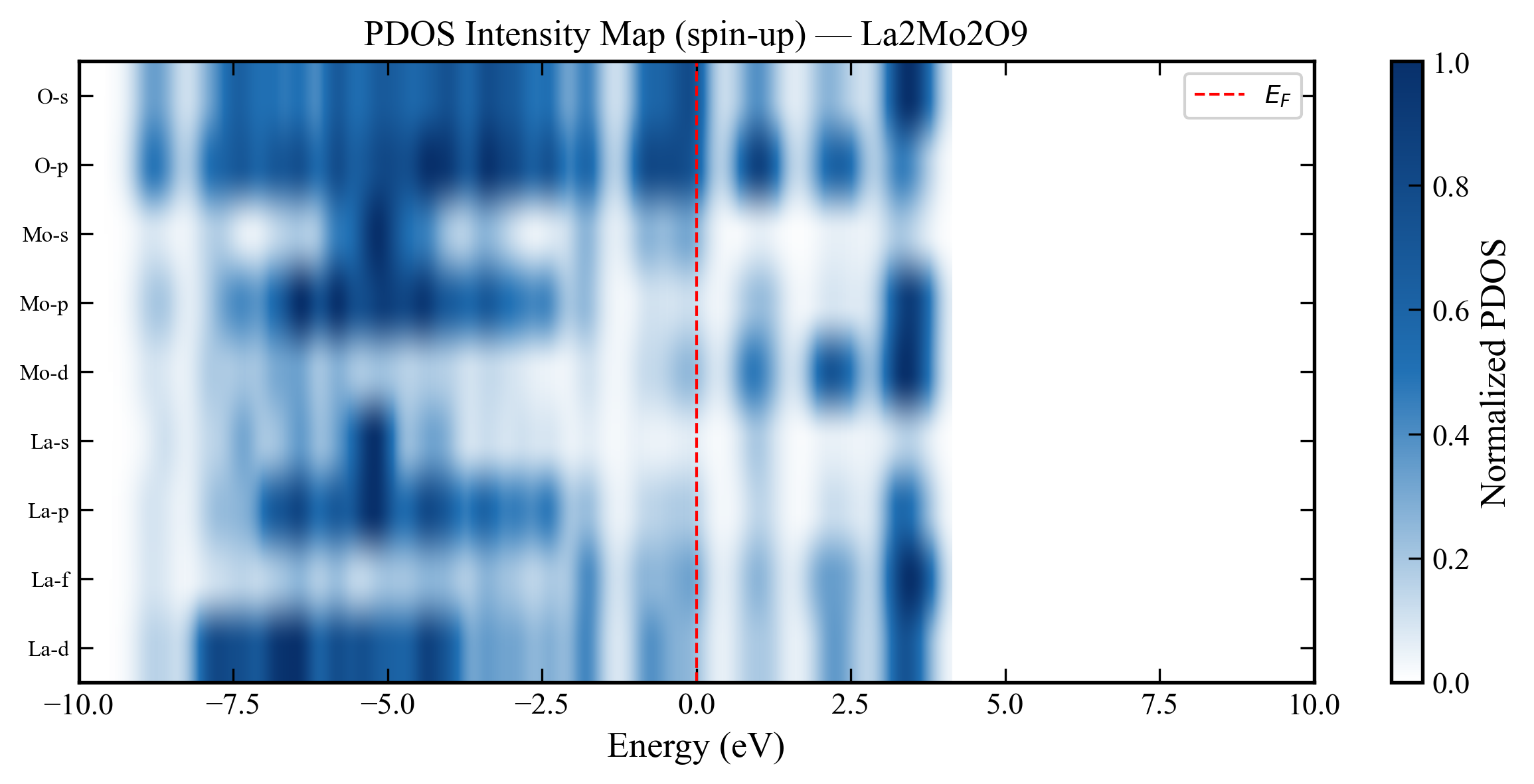}
	\caption{Row-normalised PDOS intensity map (spin-up channel) of LMO, showing the relative energy distribution of each orbital-resolved PDOS channel. The red dashed line marks $E_F$.}
	\label{fig:lmo-heatmap}
\end{figure}

The row-normalised PDOS intensity map of Fig.~\ref{fig:lmo-heatmap} makes the same point visually: the O-$s$/O-$p$ rows carry a broad, continuous band of intensity spanning nearly the whole occupied window, the Mo-$d$ row is dark except above $E_F$, and the La-derived rows (La-$s$, La-$p$, La-$f$, La-$d$) are diffuse and low-contrast, with no sharp feature at either band edge. No La-$f$ weight sits close to $E_F$, consistent with a nominally empty $4f^0$ shell that plays no direct role in the frontier states of the undoped compound.

\subsubsection*{LMOG4: Gd-4$f$ exchange splitting and Mo/W-derived redistribution}

\begin{figure}[!htbp]
	\centering
	\includegraphics[width=0.78\textwidth]{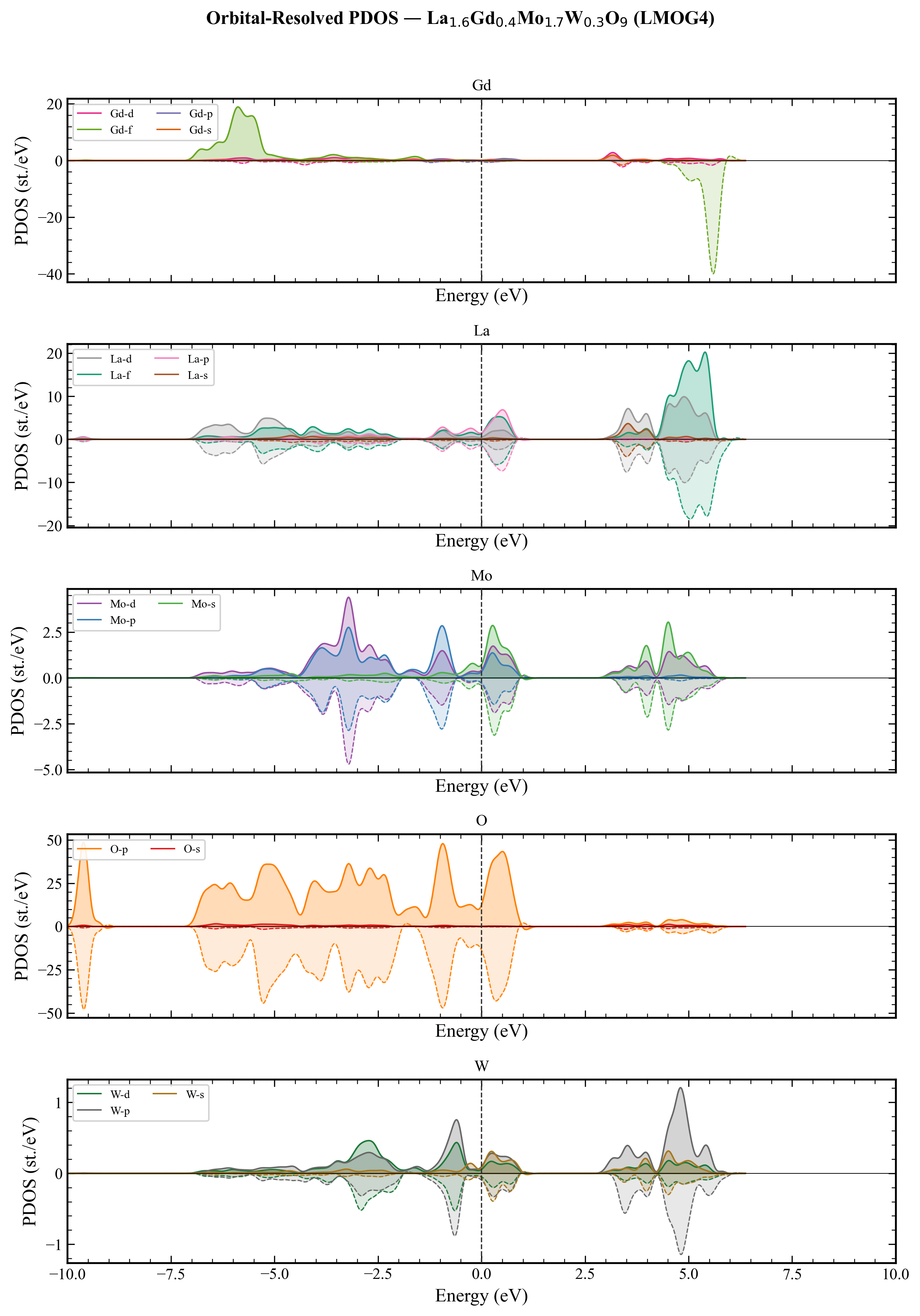}
	\caption{Orbital-resolved, spin-decomposed projected density of states of LMOG4 (La$_{1.6}$Gd$_{0.4}$Mo$_{1.7}$W$_{0.3}$O$_9$), separated by element (Gd, La, Mo, O, W). Solid curves denote the spin-up channel and dashed curves the spin-down channel (plotted as negative PDOS for clarity); the vertical dashed line marks $E_F$. Note the different vertical scale of the Gd panel relative to the other elements.}
	\label{fig:lmog4-pdos-stacked}
\end{figure}

Co-doping with Gd on the La site and W on the Mo site changes this picture substantially. The most conspicuous feature of Fig.~\ref{fig:lmog4-pdos-stacked} is the Gd panel: a single, intense majority-spin peak at $E-E_F \approx -5.5$ to $-6$~eV, reaching $\sim$20 states/eV, well above any other channel in that window, and an equally intense minority-spin peak of opposite sign near $+5$ to $+5.5$~eV ($\sim$40 states/eV). The occupied majority-spin peak coincides with two of the principal total-DOS maxima of LMOG4, at $E-E_F = -5.992$ and $-6.422$~eV. These two features are, therefore, largely Gd-4$f$ in origin rather than purely oxygen-derived, unlike the corresponding energy window in undoped LMO. The $\sim$10--11~eV majority-minority splitting is the expected electronic-structure signature of the half-filled $4f^7$ ($^8S_{7/2}$) configuration of Gd$^{3+}$, and it only emerges here because an on-site Hubbard correction of $U=6.0$~eV was applied to the Gd-4$f$ manifold (Sec.~\ref{sec:dft-implementation}). Without it, semi-local GGA is known to place partially filled $4f$ states too close to $E_F$, an artefact of uncorrected self-interaction error \cite{Cococcioni2005,Anisimov1991}. Even with the correction, a small residual Gd-derived shoulder (Gd-$p$/Gd-$d$, $<1$ states/eV) persists within $\pm 1$~eV of $E_F$ (Fig.~\ref{fig:lmog4-pdos-stacked}, Gd panel). This is reflected in the near-$E_F$ decomposition of Table~\ref{tab:pdos-ef}, and matches the residual near-gap $4f$ weight typically reported for DFT+U treatments of Gd-based oxides \cite{Cococcioni2005}.

A second change concerns La. In LMOG4 the La-$f$ and La-$d$ channels carry markedly more conduction-band weight ($\sim$20 states/eV near $+4$ to $+6$~eV) than in LMO ($\sim$1.5 states/eV near $+3.5$~eV; compare the La panels of Figs.~\ref{fig:lmo-pdos-stacked} and \ref{fig:lmog4-pdos-stacked}). La$^{3+}$ has no partially occupied $f$ states, so this enhancement is best read as hybridisation of nominally La-projected orbitals with the nearby, energetically close, unoccupied Gd-4$f$ manifold, rather than a genuine La-derived feature. This is an artefact worth flagging when comparing PBE-GGA rare-earth PDOS across doped and undoped cells \cite{Cococcioni2005}.

The Mo panel of Fig.~\ref{fig:lmog4-pdos-stacked} looks quite different from that of LMO. Conduction-band Mo-$d$ intensity is suppressed and redistributed: the tallest LMOG4 Mo-$d$ conduction feature reaches only $\sim$4 states/eV, about a fifth of the corresponding LMO peak, while additional Mo-$d$/Mo-$p$ weight appears within $\pm 1$~eV of $E_F$ with no direct counterpart in LMO. This redistribution tracks the larger Mo-4$d$ crystal-field splitting extracted for LMOG4 in Sec.~\ref{sec:dft-implementation} ($\Delta_{\mathrm{CF}} \approx 3.57$~eV, against $1.18$~eV in LMO), and reflects both the reduced Mo stoichiometry (1.7 versus 2.0 per formula unit) and the modified octahedral field imposed by neighbouring W$^{6+}$. The W panel is weaker but not negligible, with an occupied, oxygen-hybridised component between $-3$ and $-0.5$~eV and an unoccupied component peaking near $+4.5$ to $+5$~eV, systematically higher than the Mo-4$d$ conduction feature. This is consistent with the more extended, higher-lying character expected of $5d$ relative to $4d$ orbitals for formally isovalent W$^{6+}$/Mo$^{6+}$ centres. The O-2$p$ channel is still the dominant valence-band contributor in LMOG4, but its weight is lower and more finely structured (several resolved peaks between $-7.5$ and $-1$~eV, Fig.~\ref{fig:lmog4-pdos-stacked}), matching the $\sim$30\% reduction in integrated O-2$p$ valence weight reported in Sec.~\ref{sec:dft-implementation}.

\begin{figure}[!htbp]
	\centering
	\includegraphics[width=0.70\textwidth]{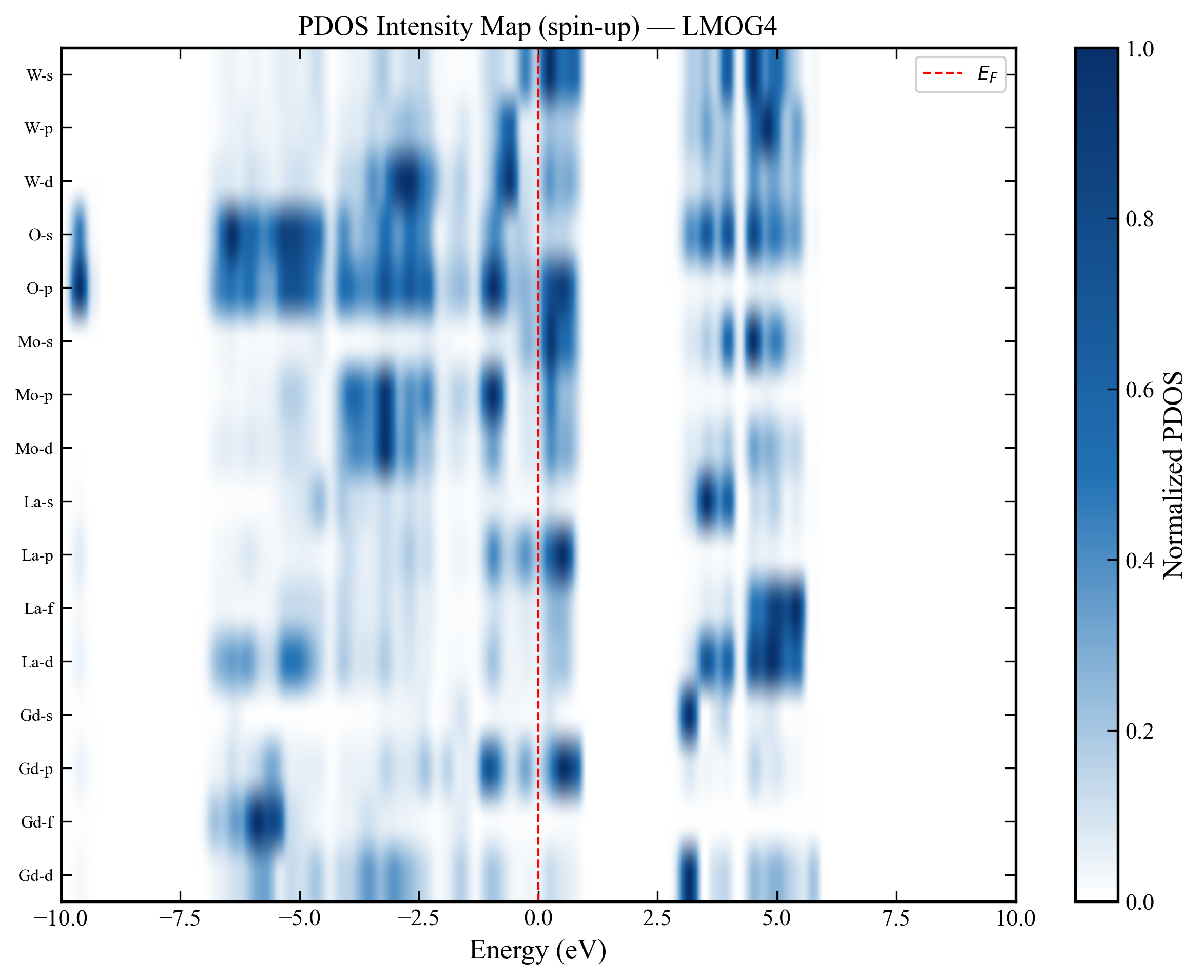}
	\caption{Row-normalised PDOS intensity map (spin-up channel) of LMOG4, showing the relative energy distribution of each orbital-resolved PDOS channel. The red dashed line marks $E_F$.}
	\label{fig:lmog4-heatmap}
\end{figure}

The intensity map of Fig.~\ref{fig:lmog4-heatmap} summarises these trends. The Gd-$f$ row shows the two darkest, most spatially confined features in the map, one below and one above $E_F$. The La-$f$/La-$d$ rows are dark only in the conduction band. The O-$p$/O-$s$ rows keep broad occupied-band intensity similar to LMO, though truncated at somewhat lower energies. The W-derived rows show faint, diffuse intensity on both sides of $E_F$, without a sharp isolated peak, distinguishing their delocalised character from the localised Gd-4$f$ states.

\subsubsection*{Orbital decomposition at the Fermi level}

To see how doping redistributes spectral weight precisely at $E_F$, where it matters most for conduction, the orbital contributions were integrated over a narrow window $\Delta E = 0.5$~eV about $E_F$,

\begin{equation}
	w_{A,l} = \frac{1}{2}\left[\rho^{\uparrow}_{A,l}(E)+\rho^{\downarrow}_{A,l}(E)\right]\bigg|_{|E-E_F|<0.5~\mathrm{eV}}, \qquad
	c_{A,l} = \frac{w_{A,l}}{\sum_{A',l'} w_{A',l'}} \times 100\%,
	\label{eq:pdos-frac}
\end{equation}

\noindent with results in Table~\ref{tab:pdos-ef}. In LMO, O-2$p$ alone accounts for 78.1\% of the near-$E_F$ weight, Mo-4$d$ a further 13.8\%, and every other channel stays below 4\%. In LMOG4, the O-2$p$ fraction drops to 66.2\%, while La-$p$ (9.5\%), La-$f$ (8.2\%) and Mo-$s$ (4.0\%) each pick up a non-trivial share, and Gd itself contributes a modest 1.8\% (Gd-$p$ + Gd-$f$ combined) despite the Hubbard correction pushing most of its weight away from $E_F$. Oxygen loses ground to a broader mix of La-, Mo- and Gd-derived channels. This is the PDOS-level counterpart of the $\sim$83\% rise in the O-metal hybridisation index and the $-29.8\%$ drop in integrated O-2$p$ valence weight already reported in Sec.~\ref{sec:dft-implementation}, and it points to co-doping diversifying, rather than simply diluting, the orbital origin of the states available for transport near $E_F$.

\vspace{-0.4em}
\begin{table}[!htbp]
	\centering
	\caption{Orbital contributions to the projected density of states within $\pm 0.5$~eV of $E_F$, Eq.~\eqref{eq:pdos-frac}, for LMO and LMOG4. Only channels contributing more than 1\% in at least one material are listed.}
	\label{tab:pdos-ef}
	\begin{tabular}{lcc}
		\toprule
		Orbital channel & LMO $c_{A,l}$ (\%) & LMOG4 $c_{A,l}$ (\%) \\
		\midrule
		O-2$p$   & 78.1 & 66.2 \\
		Mo-4$d$  & 13.8 & 2.3 \\
		O-2$s$   & 3.5  & 0.6 \\
		La-5$d$  & 1.4  & 3.1 \\
		La-4$f$  & 1.3  & 8.2 \\
		La-6$p$  & 1.2  & 9.5 \\
		Mo-5$s$  & 0.5  & 4.0 \\
		Mo-5$p$  & 0.1  & 1.8 \\
		Gd-6$p$  & --   & 1.0 \\
		Gd-4$f$  & --   & 0.8 \\
		\bottomrule
	\end{tabular}
\end{table}
\vspace{-0.4em}

\subsubsection*{Comparison with the wider LAMOX PDOS literature}

The O-2$p$/Mo-4$d$ frontier-orbital picture obtained here for LMO agrees, at the qualitative level, with the only directly comparable PDOS study of pristine La$_2$Mo$_2$O$_9$: it, too, assigns the valence-band edge to O-2$p$ and the conduction-band edge to Mo-4$d$, with La states pushed to higher energy away from both edges \cite{Colmont2020Origin}. Low-level Bi substitution for La leaves this pattern essentially unperturbed \cite{Colmont2020Origin}, in line with the general expectation, drawn from first-principles studies of oxide-ion conductors more broadly, that isovalent A-site substitution mainly reorganises the local oxygen-vacancy landscape and leaves the frontier states alone \cite{MunozGarcia2014}. The present results extend this to a heterovalent-radius, dual-site (La/Gd and Mo/W) substitution, and the PDOS response is much stronger than for simple Bi-for-La doping: a resolvable, strongly exchange-split Gd-4$f$ manifold, a substantially redistributed Mo-4$d$ conduction band, and a measurable dilution of O-2$p$ character at $E_F$. A recent PBE+D3 study of transition-metal-doped La$_2$Mo$_2$O$_9$ reports that the computed band gap widens with increasing Er content \cite{Gaur2024Study}; our PBE+U calculations instead keep a finite, metallic PDOS at $E_F$ for both LMO and LMOG4. The difference likely traces to the distinct electronic character of the Gd$^{3+}$ ($4f^7$) dopant relative to other rare-earth/transition-metal substituents, and to the explicit Hubbard treatment of Gd-4$f$ adopted here, rather than to any disagreement in the underlying O-2$p$/Mo-4$d$ framework shared by both studies.

\subsection{Correlation with experimental and DFT output}
\label{sec:correlation}

The LMO and LMOG4 supercells treated in Secs.~\ref{sec:building-supercell}--\ref{sec:pdos} correspond to the two compositional end-members ($x=0$ and $x=0.4$) of the experimental series La$_{2-x}$Gd$_x$Mo$_{1.7}$W$_{0.3}$O$_{9-\delta}$ reported by Shiwankar \textit{et al.} \cite{Shiwankar2025}, who synthesised the full series ($x=0,\,0.1,\,0.2,\,0.3,\,0.4$; LMO, LMOG1, LMOG2, LMOG3, LMOG4 in the naming convention used here) by the Pechini route, and characterised it by laboratory and angle-dispersive X-ray diffraction (Rietveld refinement, \texttt{FullProf}), Raman spectroscopy, and complex impedance spectroscopy. Since that study and the present calculations describe the same nominal chemistry, it is a direct, composition-matched benchmark for the structural inputs and electronic-structure trends of Secs.~\ref{sec:building-supercell}--\ref{sec:pdos}, and a point of entry into the wider mechanistic literature on doped La$_2$Mo$_2$O$_9$ (LAMOX).

\subsubsection*{Structural correlation}

Rietveld refinement of the experimental series gives a cubic $P2_13$ structure for every co-doped composition, against a monoclinic ($P2_1$) parent compound at room temperature. Angle-dispersive XRD confirms that Gd/W co-doping suppresses the order--disorder ($\alpha\to\beta$) transition responsible for this monoclinic distortion, so the cubic, oxide-ion-conducting $\beta$-type framework survives down to room temperature once Gd and W are present \cite{Shiwankar2025}. This is the structural picture already assumed in the DFT calculations of Sec.~\ref{sec:dft-implementation}: the LMO cell was built in the rhombohedral/pseudo-cubic setting appropriate to the higher-symmetry LAMOX parent lattice, and the LMOG4 cell as a triclinic supercell derived from the same cubic-type metric.

The refined cubic lattice parameter contracts with Gd content, from $a = 7.1478$~\AA\ ($V = 365.19$~\AA$^3$) for LMO to $a = 7.1300$~\AA\ ($V = 362.50$~\AA$^3$) for LMOG4: $-0.25\%$ in $a$, $-0.74\%$ in cell volume \cite{Shiwankar2025}, attributed to the smaller effective ionic radius of Gd$^{3+}$ relative to La$^{3+}$ at eight-fold coordination \cite{Shannon1976}. Table~\ref{tab:structure-series} lists the refined lattice parameters, cell volumes and Rietveld agreement factors for the full series. The DFT-input lattice vectors used to define the LMO and LMOG4 supercells (Sec.~\ref{sec:dft-implementation}) correspond to cubic-equivalent metrics of $7.1478$ and $7.1295$~\AA, matching the experimental end-members to within $0.001$~\AA\ ($<0.02\%$). The DFT-relaxed geometries are, on this evidence, a faithful representation of the two synthesised compositions, not an idealised or arbitrary choice.

\begin{table}[!htbp]
	\centering
	\caption{Rietveld-refined structural parameters of the La$_{2-x}$Gd$_x$Mo$_{1.7}$W$_{0.3}$O$_{9-\delta}$ series, after Shiwankar \textit{et al.} \cite{Shiwankar2025}. $\chi^2$, $R_p$ and $R_{wp}$ are the Rietveld goodness-of-fit and profile agreement factors.}
	\label{tab:structure-series}
	\begin{tabular}{lccccc}
		\toprule
		Sample & $x$ (Gd) & $a$ (\AA) & $V$ (\AA$^3$) & $R_p$ (\%) & $R_{wp}$ (\%) \\
		\midrule
		LMO   & 0.0 & 7.1478 & 365.19 & 21.9 & 19.8 \\
		LMOG1 & 0.1 & 7.1473 & 365.11 & 15.2 & 14.6 \\
		LMOG2 & 0.2 & 7.1416 & 364.25 & 15.9 & 14.2 \\
		LMOG3 & 0.3 & 7.1363 & 363.43 & 14.5 & 12.9 \\
		LMOG4 & 0.4 & 7.1300 & 362.50 & 24.7 & 20.5 \\
		\bottomrule
	\end{tabular}
\end{table}

\subsubsection*{Vibrational and bonding correlation}

Raman spectroscopy of the same series shows the symmetric Mo--O stretching mode $\nu_3(\mathrm{MoO_4})$ red-shifting from $\sim$793~cm$^{-1}$ in LMO to $\sim$777~cm$^{-1}$ from LMOG2 onward, while its relative intensity falls sharply on first doping (to $\sim$23\% of the LMO intensity in LMOG1), partly recovers in LMOG2--LMOG3, and drops again to $\sim$57\% in LMOG4 \cite{Shiwankar2025}. A Raman red-shift of a metal--oxygen stretching mode normally signals a weaker bond force constant, i.e., a softer Mo--O bond \cite{Colmont2020Origin}. This tracks the DFT covalency indicators of Sec.~\ref{sec:dft-implementation}: the integrated O-2$p$ valence weight falls $-29.8\%$ from LMO to LMOG4, the metal--oxygen hybridisation index rises $+83.5\%$, and the Mo-4$d$ crystal-field splitting more than triples ($\Delta_{\mathrm{CF}}$: 1.18~eV~$\to$~3.57~eV). Together, these metrics describe an M--O bonding environment that becomes electronically softer and more strongly redistributed on Gd/W co-doping, in the same direction as the observed vibrational softening. The two probes should not be expected to agree quantitatively: Raman scattering samples the vibrational, finite-temperature response of the lattice, while the PDOS-based hybridisation index is a $T=0$~K electronic-structure quantity. The correlation drawn here is one of consistent trend, not a direct numerical mapping.

\subsubsection*{Transport correlation}

Complex impedance spectroscopy shows that bulk ionic conductivity in the series does not vary monotonically with Gd content. At 650~$^{\circ}$C, $\sigma$ rises from $8.99\times10^{-3}$~S~cm$^{-1}$ in LMO to a maximum of $2.07\times10^{-2}$~S~cm$^{-1}$ in LMOG3 ($x=0.3$), then falls sharply to $1.54\times10^{-3}$~S~cm$^{-1}$ in LMOG4 ($x=0.4$), roughly an order of magnitude below the LMOG3 optimum \cite{Shiwankar2025}. The Arrhenius activation energy, $\sigma T = \sigma_0 \exp\!\left(-E_a/k_BT\right)$, follows the mirror trend: it drops from $E_a = 1.00$~eV in LMO at high temperature (with a lower-temperature value of $1.90$~eV reflecting the extra contribution of the order--disorder transition) to $1.05$~eV in LMOG1, then climbs through $1.20$~eV (LMOG2) and $1.22$~eV (LMOG3) to $1.46$~eV in LMOG4 \cite{Shiwankar2025}. Above the transition region, several LAMOX-type compositions depart from simple Arrhenius behaviour and are better described by a Vogel--Tammann--Fulcher (VTF) expression,

\begin{equation}
	\sigma T = \sigma_0 \exp\!\left[-\frac{E_a}{k_B\left(T-T_0\right)}\right],
	\label{eq:vtf}
\end{equation}

\noindent where $T_0$ is an empirical reference temperature below the nominal transport onset. This crossover has been linked to long-range oxide-ion migration coupling to increasingly disordered, dynamically fluctuating oxygen sublattices in the high-temperature LAMOX framework \cite{Malavasi2007Nature,Lacorre2006}.

Table~\ref{tab:transport-series} collects the conductivity and activation-energy data for the full series, alongside two DFT indicators computed at the two end-members in this work: the O-2$p$/metal hybridisation index $H$, and the integrated spin asymmetry $\Delta N = N_{\uparrow}-N_{\downarrow}$ of the occupied band (Sec.~\ref{sec:dft-implementation}). A two-point DFT comparison cannot, on its own, explain the non-monotonic conductivity trend, since LMO and LMOG4 are the only compositions treated computationally here. It is, however, directly consistent with the conductivity collapse observed at LMOG4. That composition shows (i) the largest Mo-4$d$ crystal-field splitting implied by the DFT trend, $+202\%$ relative to LMO, which should locally rigidify the Mo/W--O polyhedral network rather than promote the flexibility associated with facile oxide-ion hopping \cite{Corbel2011Structural,Lacorre2006}, and (ii) a resolvable, strongly exchange-split Gd-4$f$ manifold with residual spectral weight near $E_F$ (Sec.~\ref{sec:pdos}), consistent with a more strongly localised, magnetically active dopant sublattice at the highest Gd content. Both points match the interpretation offered for the analogous Eu-doped series La$_{2-x}$Eu$_x$Mo$_2$O$_9$: high dopant content can stabilise the cubic framework structurally while introducing local distortions or defect-trapping tendencies unfavourable for oxide-ion transport \cite{Corbel2007Topological}.

\begin{table}[!htbp]
	\centering
	\caption{Bulk ionic conductivity at 650~$^{\circ}$C and Arrhenius activation energy across the La$_{2-x}$Gd$_x$Mo$_{1.7}$W$_{0.3}$O$_{9-\delta}$ series, after Shiwankar \textit{et al.} \cite{Shiwankar2025}, alongside DFT electronic-structure indicators computed in this work for the two end-members (Sec.~\ref{sec:dft-implementation}).}
	\label{tab:transport-series}
	\begin{tabular}{lcccc}
		\toprule
		Sample & $\sigma_{650^{\circ}\mathrm{C}}$ (S cm$^{-1}$) & $E_a$ (eV) & $H$ (DFT) & $\Delta N$ (DFT, states) \\
		\midrule
		LMO   & $8.99\times10^{-3}$ & 1.90 / 1.00 & 0.627 & $\approx 0.01$ \\
		LMOG1 & $2.97\times10^{-3}$ & 1.05 & --   & -- \\
		LMOG2 & $1.23\times10^{-2}$ & 1.20 & --   & -- \\
		LMOG3 & $2.07\times10^{-2}$ & 1.22 & --   & -- \\
		LMOG4 & $1.54\times10^{-3}$ & 1.46 & 1.151 & 17.1 \\
		\bottomrule
	\end{tabular}
\end{table}

\subsubsection*{Dynamical disorder and diffusion pathway: comparison with a Dy, W co-doped LMX system}

The non-monotonic doping response of Table~\ref{tab:transport-series}, a conductivity optimum at intermediate Gd content and collapse at the highest content studied, is not unique to the Gd/W series. Nayyer \textit{et al.} report closely analogous behaviour in a chemically distinct but structurally equivalent system, La$_{2-x}$Dy$_x$Mo$_{1.7}$W$_{0.3}$O$_9$ ($0.1 \le x \le 0.5$), where Dy$^{3+}$ rather than Gd$^{3+}$ occupies the La site alongside the same W$^{6+}$-for-Mo$^{6+}$ substitution \cite{Nayyer2020Dyn}. Their XRD data show the same disappearance of the (211), (200) and (231) peak splitting on co-doping that Shiwankar \textit{et al.} use to confirm the suppressed $\alpha\to\beta$ transition in the Gd/W series \cite{Shiwankar2025,Nayyer2020Dyn}, so the structural correlation drawn above for LMO/LMOG4 appears to hold for rare-earth co-doping of La$_2$Mo$_2$O$_9$ more generally, and is not an artefact specific to the Gd$^{3+}$ ionic radius.

Because that study probes the sample dynamically, using high-temperature Raman, dielectric relaxation, and complex-electric-modulus spectroscopy, it offers a mechanistic layer that the present, static, $T=0$~K PDOS calculations cannot reach on their own. Two results stand out. First, the rate of change of the Mo--O Raman intensity with temperature, $dI/dT$, is non-uniform in every co-doped composition, and this is taken as the direct spectroscopic signature of a dynamically disordered oxygen-vacancy sublattice, distinct from the long-range-ordered, static disorder of pure LMO \cite{Nayyer2020Dyn}. Second, dielectric loss spectra resolve a single relaxation peak near 400~$^{\circ}$C in pure LMX, attributed to a single O(1) exchange pathway through orderly arranged vacancies, against two relaxation peaks between 350 and 550~$^{\circ}$C in every Dy,W co-doped composition, assigned to two independent O(1)$\to$O(2) and O(1)$\to$O(3) ion-vacancy exchange routes opened up once oxygen vacancies redistribute onto the partially occupied O2 and O3 sites of the cubic $\beta$ phase \cite{Nayyer2020Dyn}. The imaginary part of the electric modulus, $M''(\omega)$, was fit for every composition with the Bergman form of the Kohlrausch--Williams--Watts function,

\begin{equation}
	M''(\omega) = \frac{M''_{\max}}{(1-\beta) + \dfrac{\beta}{1+\beta}\left[\beta\left(\dfrac{\omega_{\max}}{\omega}\right) + \left(\dfrac{\omega}{\omega_{\max}}\right)^{\beta}\right]},
	\label{eq:kww-modulus}
\end{equation}

\noindent with the stretching exponent $\beta<1$ recovered in every case, confirming non-Debye relaxation dynamics throughout the series \cite{Nayyer2020Dyn}.

The activation energy extracted from the modulus peak frequency, $f_{\max}=f_0\exp(-E_a/k_BT)$, is 1.49~eV below and 1.32~eV above the $\sim$560~$^{\circ}$C transition in pure LMX, and it falls with light Dy doping ($x=0.1$--$0.3$) before rising again at the two heaviest compositions ($x=0.4$--$0.5$) \cite{Nayyer2020Dyn}. The impedance-derived Arrhenius activation energy shows the same qualitative shape, from 0.98~eV at $x=0.3$ up to 1.20~eV at $x=0.5$ \cite{Nayyer2020Dyn}, mirroring the intermediate-composition optimum and high-doping penalty already seen for $E_a$ in the Gd/W series (Table~\ref{tab:transport-series}). A single-slope Arrhenius law is recovered across the full 400--700~$^{\circ}$C IT-SOFC window for every Dy,W co-doped composition, in place of the discontinuous jump in conductivity that pure LMX shows at its $\alpha\to\beta$ transition \cite{Nayyer2020Dyn}. That two chemically different lanthanide dopants (Dy$^{3+}$, Gd$^{3+}$) paired with the same W$^{6+}$ co-dopant give the same qualitative outcome, phase-transition suppression, a two-pathway vacancy exchange mechanism, and a non-monotonic activation-energy trend with an intermediate-composition optimum, suggests that these are general features of heavy dual-site LAMOX substitution rather than a Gd-specific coincidence, and gives an independent, dynamic-measurement basis for the DFT-derived hybridisation and crystal-field trends of Secs.~\ref{sec:dft-implementation}--\ref{sec:pdos}.

\subsubsection*{Reconciling static DFT with dynamic ionic transport}

An important caveat applies to every correlation drawn above: the present calculations are static, $T=0$~K, $\Gamma$-point electronic-structure calculations, while ionic conductivity is an intrinsically dynamic, finite-temperature transport property. Pair-distribution-function analysis of La$_2$Mo$_2$O$_9$ shows that the local structure of the conducting cubic $\beta$ phase is essentially the same as that of the low-temperature monoclinic $\alpha$ phase, so the $\alpha\to\beta$ transition is better read as an order--disorder transition of the oxygen sublattice than as a reconstructive change in the average cation framework \cite{Malavasi2007Nature}. Quasi-elastic neutron scattering combined with \textit{ab initio} molecular dynamics shows that oxide-ion migration in La$_2$Mo$_2$O$_9$ proceeds by discrete jumps within and between Mo coordination polyhedra, with a nanosecond-timescale activation energy of $0.61(5)$~eV \cite{Peet2017Direct}, markedly lower than the macroscopic Arrhenius values of Table~\ref{tab:transport-series}. The difference reflects extra contributions to impedance-derived $E_a$, such as grain-boundary and long-range percolation effects, absent from a local, single-jump picture, and it is broadly consistent with the two distinct vacancy-hopping pathways resolved by dielectric relaxation in the Dy,W-codoped system above \cite{Nayyer2020Dyn}.

The static PDOS reported in Sec.~\ref{sec:pdos} identifies where electronic charge sits (mostly O-2$p$ in the valence band, with a growing La/Mo/Gd admixture at $E_F$ on doping) and how strongly the metal and oxygen orbitals hybridise, but it does not by itself fix the oxide-ion migration barrier. That barrier is set by the dynamic flexibility of the Mo/W-centred polyhedral network and the associated oxygen-vacancy energetics \cite{MunozGarcia2014,Lacorre2006}, which lie outside the scope of the present ground-state calculations, and are exactly what the Raman, dielectric and modulus measurements of Nayyer \textit{et al.} probe directly \cite{Nayyer2020Dyn}. Within this limitation, the DFT-derived covalency and crystal-field trends of Secs.~\ref{sec:dft-implementation}--\ref{sec:pdos} give a physically consistent, composition-resolved electronic-structure counterpart to the structural and transport behaviour measured for the same LMO/LMOG4 end-members by Shiwankar \textit{et al.} \cite{Shiwankar2025}, and to the dynamical-disorder mechanism established, on a related dual-doped LMX system, by Nayyer \textit{et al.} \cite{Nayyer2020Dyn}.

\section{CONCLUSIONS}
\label{sec:conclusions}

We have presented a first-principles electronic-structure study of pristine La$_2$Mo$_2$O$_9$ (LMO) and its Gd/W co-doped derivative La$_{1.6}$Gd$_{0.4}$Mo$_{1.7}$W$_{0.3}$O$_9$ (LMOG4), the two compositional end-members of a recently reported co-doped series, built on symmetry-optimised 144-atom supercells and treated within spin-polarised DFT+U with an on-site Hubbard correction on the localised Gd-$4f$ manifold. The projected density of states shows that pristine LMO is an O-2$p$/Mo-4$d$ charge-transfer oxide in which La contributes negligibly at either band edge. Gd/W co-doping reorganises this picture substantially: a strongly exchange-split Gd-4$f$ doublet emerges, resolvable only once the Hubbard correction is included; the Mo-4$d$ conduction band is redistributed, with its crystal-field splitting more than tripling; and the O-2$p$ contribution at the Fermi level falls from 78\% to 66\% as La-, Mo- and Gd-derived channels acquire a resolvable share. These electronic-structure trends track, at the microscopic level, the lattice contraction, Mo--O Raman softening and non-monotonic ionic conductivity reported experimentally for the same series, and identify the pronounced Mo-4$d$ crystal-field stiffening and the onset of localised Gd magnetism as plausible electronic correlates of the conductivity collapse observed at the highest Gd content, rather than structural instability alone.

These findings are subject to the limitations inherent in a static, zero-temperature, $\Gamma$-point treatment of only two compositions, and to the well-known tendency of semi-local exchange-correlation functionals to underestimate the band gap of $d$-electron oxides even with a Hubbard correction confined to the rare-earth $4f$ shell. The supercells constructed here were selected specifically to support extraction of the underlying Mo--Mo and Gd-mediated exchange constants, and doing so, alongside extending the present treatment to the full compositional series and to finite-temperature estimates of oxygen migration, is the natural next step. More broadly, the results indicate that rare-earth/transition-metal co-doping of LAMOX electrolytes cannot be understood through structural stabilisation of the cubic framework alone: the accompanying redistribution of orbital character at the Fermi level, and the emergence of a genuinely new, localised magnetic sublattice, are integral to how such co-doping ultimately shapes ionic transport.
\section{ACKNOWLEDGEMENT}
\label{sec:acknowledgement}
The authors thank the Department of Physics, Rashtrasant Tukadoji Maharaj Nagpur University, for institutional and computational support extended to the Advanced Materials Research Laboratory during the course of this work. The authors acknowledge the financial support from UGC-DAE CSR through a Collaborative Research Scheme (CRS) project number CRS/2023-24/01/1023.

\section{DATA AVAILABILITY}
\label{sec:data_availability}
The Quantum ESPRESSO input files for LMO and LMOG4, the SuperHEX supercell-generation configuration files and the corresponding structure-selection data (\texttt{struct\_analysis.csv}) reported in this work are openly available in a GitHub repository at \url{https://github.com/amoghlanjewar/lmo-lmog4-dft.git}. Raw Quantum ESPRESSO wavefunction and charge-density files, being large binary outputs, are available from the corresponding author upon reasonable request.

\renewcommand{\refname}{References}

\bibliographystyle{apsrev4-2}
\bibliography{Project}

\end{document}